\documentclass[sigconf]{acmart}

\AtBeginDocument{%
  }

\usepackage{amsmath}
\usepackage{bm}
\usepackage{mathtools}
\usepackage{algorithm}
\usepackage{algpseudocode}
\usepackage{xspace}
\usepackage{multirow}
\usepackage{array}
\usepackage{tabularx}
\usepackage{booktabs}
\usepackage{threeparttable}
\usepackage{graphicx}
\usepackage{enumitem}
\usepackage{placeins}

\usepackage{subfigure}

\newcommand{\method}{\textsc{DyGen}\xspace}

\begin{document}

\title{Generative Learning on Dynamic Text-Attributed Graphs}

\author{Zhenyu Yang}
\affiliation{
  \institution{Xiamen University}
  \city{Xiamen}
  \country{China}
}
\email{27720251153998@stu.xmu.edu.cn}

\author{Yuhong Chen}
\affiliation{
  \institution{Xiamen University}
  \city{Xiamen}
  \country{China}
}
\email{yhchen2320@163.com}

\author{Binli Luo}
\affiliation{
  \institution{Central South University}
  \city{Changsha}
  \country{China}
}
\email{lblhandsome@gmail.com}

\author{Xiaodong Li}
\affiliation{
  \institution{Xiamen University}
  \city{Xiamen}
  \country{China}
}
\email{xdli@xmu.edu.cn}

\author{Xiaolin Hu}
\affiliation{
  \institution{Xiamen University}
  \city{Xiamen}
  \country{China}
}
\email{xiaolinhu@xmu.edu.cn}

\author{Hui Li}
\affiliation{
  \institution{Xiamen University}
  \city{Xiamen}
  \country{China}
}
\email{hui@xmu.edu.cn}

\author{Jintang Li}
\affiliation{
  \institution{Xiamen University}
  \city{Xiamen}
  \country{China}
}
\email{edisonlee@xmu.edu.cn}
% WSDM submissions must be anonymous. Restore the actual author block only
% after acceptance.
% \author{Anonymous Author(s)}
\renewcommand{\shortauthors}{Anonymous Author(s)}

\begin{abstract}
Dynamic text-attributed graphs capture evolving interactions among entities with rich textual information, but future-link prediction remains difficult when nodes have little interaction history or appear as new targets. We propose \method, a generative framework that represents nodes with compact semantic identifiers (SIDs) learned from text and autoregressively predicts a target SID conditioned on the source SID, recent interaction events, and relative time gaps. A lightweight structural calibration further incorporates repetition, co-occurrence, and popularity. At inference time, \method encodes each query context once and reuses SID likelihoods to rank prescribed candidates, providing a candidate-efficient prediction interface. Experiments on nine datasets show broadly competitive performance across domains and graph densities. Source-side inductive evaluations demonstrate that text-derived SIDs provide a useful representation when a test source has no observed interaction history, while two recommendation-oriented case studies evaluate controlled new-user adaptation and first-arrival ranking for inductive new items. Further analyses characterize SID uniqueness, capacity, and temporal stability. Overall, the results support combining text-derived SIDs with temporal histories for dynamic link prediction. An anonymous implementation is available at \url{https://anonymous.4open.science/status/DyGen-EC70}.
\end{abstract}




\maketitle

\section{Introduction}

Dynamic graphs capture evolving interactions in applications such as online review platforms, social networks, collaborative knowledge bases, recommender systems, and citation networks~\cite{FengWWSWH26,graphssm,EvolveGCN}. These interactions are often accompanied by rich textual attributes, including reviews, posts, profiles, item descriptions, paper abstracts, and relation descriptions~\cite{DTGB,GDGB}. Such data naturally form dynamic text-attributed graphs (DyTAGs), where temporal structure and textual attributes jointly shape dynamic graph learning~\cite{GAD,dygrasp}. As these graphs evolve, a key challenge is how to combine textual semantics with temporal interaction evidence when nodes have highly uneven interaction histories.

This challenge is particularly evident across source-history
regimes \cite{LiZLTZZ24,zheng2023decoupled}. For inductive or rarely active source nodes, structural evidence is weak or absent, making textual attributes the primary history-independent signal. A diagnostic with DyGFormer illustrates this gap: on Wikirevision, IMDB, and Weibo-Daily, its Hits@10 for sources with no training or validation interactions is only 54\%, 17\%, and 2\%, respectively, of that for sources with more than ten prior interactions. This disparity motivates a model that relies on text-derived semantics when source history is unavailable and progressively incorporates event-level temporal evidence as interactions accumulate.

\begin{figure}
    \centering
    \includegraphics[width=\linewidth]{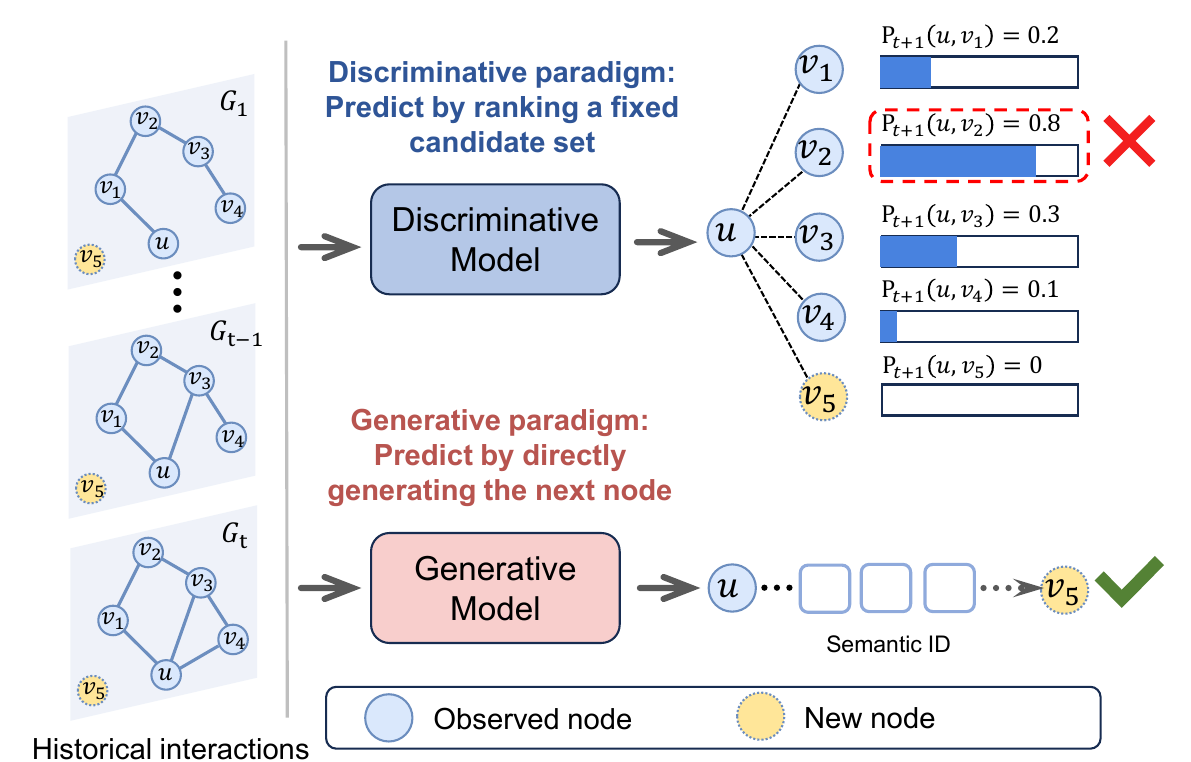}
    \caption{Comparison between discriminative and generative paradigms for dynamic graph learning. Discriminative models rank target nodes from an observed candidate set according to predicted interaction probabilities, while generative models predict the next interaction target by generating its identifier. 
    % This interface better aligns with open-world dynamic graphs, where newly emerging or sparsely observed nodes require generalization beyond fixed candidate enumeration.
    }
    \label{fig:comparison}
\end{figure}

Most dynamic graph learning methods follow a discriminative prediction paradigm. Memory-based methods maintain node states updated by interaction streams~\cite{TGN,JODIE,DyRep,sad}, while neighborhood- and sequence-based methods encode temporal neighborhoods or interaction histories~\cite{TGAT,CAWN,DyGLib,GraphMixer}. Despite architectural differences, they typically predict future interactions by scoring sampled or enumerated candidate targets. While effective when structural evidence is sufficient, such models can struggle with sources that have little or no interaction history, or with targets that have not previously appeared in the target role~\cite{moment,LKD4DyTAG}.

% Generative modeling provides a complementary prediction interface. As illustrated in Figure~\ref{fig:comparison}, discriminative models directly assign scores to candidate targets, whereas a generative formulation models the identifier of the next interaction target. Recent generative retrieval methods represent each target as a sequence of discrete semantic codes and train an autoregressive model to estimate the likelihood of its identifier~\cite{TIGER,SIDhandbook,DSI}. Residual quantization provides a practical way to convert continuous content embeddings into compact code sequences~\cite{RQ-VAE22,VQVAE}. This idea is well aligned with DyTAGs because node text can be mapped into semantic identifiers that remain available even when graph history is absent. At ranking time, prescribed candidates can be evaluated through their SID likelihoods, allowing the query-side context to be reused across candidates.

In contrast, generative modeling offers a complementary prediction interface. As illustrated in Figure~\ref{fig:comparison}, discriminative methods score candidate targets directly, whereas generative methods model the identifier of the next target. Recent generative retrieval approaches represent targets as sequences of discrete semantic codes and autoregressively estimate their likelihoods~\cite{TIGER,SIDhandbook,DSI}. Residual quantization provides an effective way to derive such identifiers from continuous content embeddings~\cite{RQ-VAE22,VQVAE}. This paradigm is particularly suitable for DyTAGs: text-derived semantic identifiers remain available even without interaction history, while candidate targets can be evaluated through their SID likelihoods using a shared query context.

% However, existing generative methods do not directly address DyTAG link prediction. Semantic-ID retrieval methods are mainly designed for relatively stable catalogs and sequential user--item interactions~\cite{TIGER,SIDhandbook}. They do not explicitly model how an inductive node acquires temporal support as new interactions arrive. Recent generative DyTAG studies instead focus on graph-level generation~\cite{GDGB,moment,dygrasp}, rather than predictive learning over individual evolving interactions. What remains missing is a generative framework that supports inductive prediction through textual semantics when a node is absent from the observed interaction graph, while progressively incorporating temporal events once interaction history becomes available.

% However, semantic-ID generative retrieval methods mainly target relatively stable recommendation catalogs and sequential user--item interactions~\cite{TIGER,SIDhandbook}. Recent DyTAG studies explore graph generation~\cite{GDGB} or predictive learning with textual semantics~\cite{moment,dygrasp}, but do not formulate future-target prediction as autoregressive generation over text-derived semantic IDs with event-level temporal conditioning. This leaves open a unified framework that starts from history-independent textual semantics for inductive nodes and progressively incorporates interaction history as structural support accumulates.

However, semantic-ID generative retrieval methods mainly target relatively stable recommendation catalogs and sequential user--item interactions~\cite{TIGER,SIDhandbook}. Recent DyTAG studies explore graph generation~\cite{GDGB} or predictive learning with textual semantics~\cite{moment,dygrasp}, but do not formulate future-target prediction as autoregressive generation over text-derived semantic IDs conditioned on event-level temporal history. This leaves open a unified generative framework that remains applicable when structural history is absent and progressively exploits temporal interactions as they accumulate.

% We propose \method, a generative semantic-ID framework for dynamic link prediction on DyTAGs. \method maps node text into compact SIDs through residual quantization and uses an encoder--decoder Transformer to predict the target SID from the source SID, recent interaction targets, and relative time gaps. The query context is encoded once and reused to rank prescribed candidates by their autoregressive SID likelihoods, providing a candidate-efficient interface while retaining candidate scoring and ranking. An inference-time structural calibration further incorporates repetition, co-occurrence, and popularity. We evaluate \method on nine datasets spanning multiple domains and graph densities. It ranks first or ties for first on three of the original eight benchmarks and among the top two on seven. Additional experiments evaluate source-side inductive prediction, controlled new-user and target-inductive new-item scenarios, statistical reliability, and SID mechanisms.

In this work, we propose \method, a generative semantic-ID framework for dynamic link prediction on DyTAGs. \method maps node text into compact SIDs through residual quantization and predicts target SIDs with an encoder--decoder Transformer conditioned on the source SID, recent interaction targets, and relative time gaps. The encoded query context is shared across prescribed candidates, whose SID likelihoods are further complemented by lightweight structural calibration. Experiments on nine datasets show broadly competitive performance across domains and graph densities, with further analyses of source- and target-side inductive generalization and SID mechanisms.

Our contributions are:
\begin{itemize}[noitemsep,topsep=3pt,leftmargin=*]
    % \item We introduce a generative predictive-learning formulation for DyTAGs that identifies future interaction targets through conditional semantic-ID generation and reuses a shared query representation across prescribed candidates.
    \item We introduce a generative formulation for predictive learning on DyTAGs, modeling future interactions through conditional generation of text-derived semantic identifiers.
    % \item We develop a unified framework that combines text-derived SIDs,
    % target-identity-aware event histories with relative temporal information,
    % and lightweight structural calibration. This design supports
    % source-inductive prediction without observed interactions and progressively
    % refines predictions as temporal evidence accumulates.
    \item We design an identity-aware temporal conditioning mechanism that encodes recent interaction targets together with relative time gaps. It progressively enriches text-based predictions with temporal evidence as interactions accumulate.
    % \item We conduct extensive experiments on nine datasets, including a graph
    % with over 3.2 million interactions. \method achieves a top-two Hits@10
    % point estimate on eight of the nine datasets and the highest
    % source-inductive Hits@10 point estimate on five of eight reported
    % benchmarks. Controlled new-user and target-inductive new-item evaluations,
    % paired statistical tests, and SID mechanism analyses further substantiate
    % its effectiveness and show that historical target identities provide
    % predictive information beyond history length and timestamps.
    \item We conduct extensive experiments on nine datasets across standard and inductive settings. The results validate the effectiveness of \method and show that historical target identities provide predictive signals beyond history length and timing.
\end{itemize}

\section{Related Work}
\label{sec:related}
We position our work along three related directions: dynamic graph learning, generative modeling with semantic IDs, and generative modeling of dynamic graphs.

\subsection{Dynamic Graph Learning}
% Dynamic graph encoders predominantly follow a discriminative formulation.
% Memory-based methods, including JODIE~\cite{JODIE}, DyRep~\cite{DyRep}, and
% TGN~\cite{TGN}, maintain per-node states updated by interaction streams.
% Sequence-based methods, including TGAT~\cite{TGAT}, CAWN~\cite{CAWN},
% GraphMixer~\cite{GraphMixer}, DTFormer~\cite{dtformer}, STHN~\cite{STHN},
% and DyGFormer~\cite{DyGLib}, aggregate recent temporal neighborhoods.
% Evaluation studies have shown that easy negative sampling can make simple
% memorization baselines unexpectedly strong~\cite{EdgeBank}; DyGLib and
% DTGB subsequently standardized evaluation for temporal and text-attributed
% graphs~\cite{DyGLib,DTGB}. TGB~\cite{TGB} and TGB~2.0~\cite{TGB2.0} provide related benchmarks for
% temporal, knowledge, and heterogeneous graphs. These
% methods generally score an enumerated candidate set. In contrast, \method
% models target SIDs autoregressively and reuses one encoded query context
% across prescribed candidates.
Dynamic graph learning predominantly adopts a discriminative prediction paradigm. Memory-based methods, such as JODIE~\cite{JODIE}, DyRep~\cite{DyRep}, and TGN~\cite{TGN}, maintain node states updated by interaction streams, while neighborhood- and sequence-based methods, including TGAT~\cite{TGAT}, CAWN~\cite{CAWN}, GraphMixer~\cite{GraphMixer}, DTFormer~\cite{dtformer}, STHN~\cite{STHN}, and DyGFormer~\cite{DyGLib}, encode recent temporal context. Prior studies have exposed evaluation pitfalls in temporal interaction prediction, particularly those arising from negative sampling~\cite{EdgeBank}. Subsequent benchmarks such as DyGLib, DTGB, TGB, and TGB~2.0~\cite{DyGLib,DTGB,TGB,TGB2.0} have promoted more standardized evaluation of dynamic graph learning. Most predictive methods evaluated under these settings score candidate targets directly. \method instead models target SIDs autoregressively while sharing the encoded query context across prescribed candidates.

\subsection{Generative Modeling with Semantic IDs}
% Generative retrieval represents target entities as discrete code sequences.
% TIGER~\cite{TIGER} learns SIDs through an RQ-VAE~\cite{RQ-VAE22} and trains
% a sequence model to predict the next target. Subsequent work studies
% alternative tokenizers, collaborative signals, and parallel decoding
% ~\cite{SIDhandbook,SIDparallel,letter,GenRet}. Recent analyses also examine
% SID drift during tokenizer optimization~\cite{DigerSid}, which differs from
% assignment changes across temporal graph snapshots. Existing methods mainly
% assume relatively stable recommendation catalogs; \method instead conditions
% SID prediction on event-level histories in a dynamic graph.
Generative retrieval represents target entities with discrete semantic identifiers (SIDs) and predicts them autoregressively. TIGER~\cite{TIGER} learns SIDs using an RQ-VAE~\cite{RQ-VAE22}, while subsequent work explores alternative tokenizers, collaborative signals, and parallel decoding~\cite{SIDhandbook,SIDparallel,letter,GenRet}. Recent studies further examine SID stability during tokenizer optimization~\cite{DigerSid}, which is distinct from temporal reassignment induced by evolving graph structure. More broadly, the growing success of SID-based generative retrieval suggests a promising direction for dynamic graph learning. To our knowledge, prior work has not combined autoregressive generation over text-derived SIDs with event-level temporal conditioning for dynamic interaction prediction.

\subsection{Generative Modeling of Dynamic Graphs}
% A separate line of work synthesizes complete dynamic graphs rather than
% predicting query-level future links. TIGGER~\cite{TIGGER} generates temporal
% random walks, DG-Gen~\cite{DG-Gen} models continuous-time edge formation,
% and VRDAG~\cite{VRDAG} and VGRNN~\cite{VGRNN} use variational formulations.
% Agent-based approaches associated with GDGB generate DyTAGs through
% simulation~\cite{GDGB}. These studies are evaluated primarily by graph-level
% distribution statistics. \method instead predicts an individual future target
% and is evaluated with per-query ranking metrics.
A separate line of work studies the generation or simulation of dynamic graph evolution. TIGGER~\cite{TIGGER} generates temporal random walks, DG-Gen~\cite{DG-Gen} models continuous-time edge formation, while VRDAG~\cite{VRDAG} and VGRNN~\cite{VGRNN} adopt variational formulations to capture evolving graph structures. Recent agent-based approaches associated with GDGB further simulate the formation of DyTAGs~\cite{GDGB}. These methods mainly aim to reproduce graph-level temporal and structural patterns, and are therefore typically evaluated using distributional statistics over generated graphs. In contrast, \method focuses on predictive learning over individual evolving interactions, estimating the next target for each query rather than synthesizing graph evolution as a whole.

\begin{figure*}[t]
\centering
\includegraphics[width=0.8\linewidth]{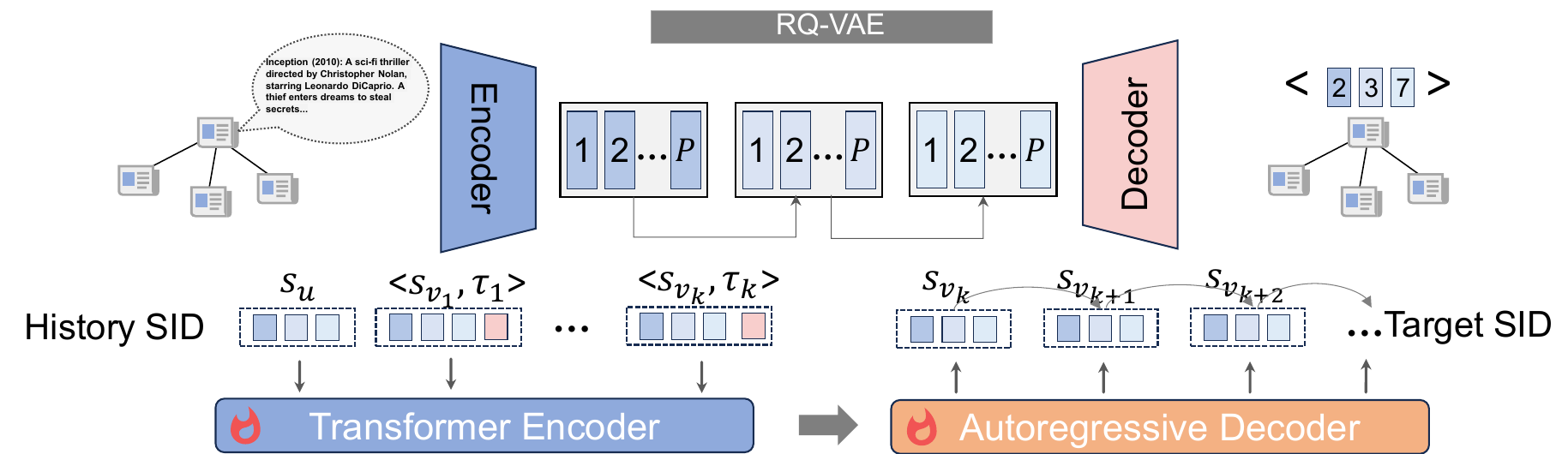}
\caption{Overall framework of \method.
\textbf{Top:} Each node's textual attribute is encoded by a frozen BERT and quantized into a length-$L$ semantic ID via a shared-codebook RQ-VAE.
\textbf{Bottom:} Input sequence construction for the Transformer pipeline, combining the source SID with historical event tokens $(\mathbf{s}_{v_k},\tau_k)$ to form the model input.}
\label{fig:framework}
\end{figure*}

\section{Preliminaries}
\label{sec:prelim}

\noindent\textbf{Definition 1 (Dynamic Text-Attributed Graph).}
A DyTAG is denoted as $G=(V,E,X)$, where $V$ is the node set,
$X=\{x_v\mid v\in V\}$ contains node textual attributes, and
$E=\{(u,v,t)\}$ contains timestamped interactions.

\noindent\textbf{Definition 2 (Dynamic Interaction Prediction).}
Let $G_{<t}=(V,E_{<t},X)$ denote the graph state available before time $t$,
where
\[
E_{<t}=\{(u',v',t')\in E\mid t'<t\}.
\]
Given a source node $u$ and query time $t$, dynamic interaction prediction
estimates the target of the next interaction:
\[
v^*=\arg\max_{v\in V_C(u,t)}
p(v\mid u,t,G_{<t},X),
\]
where $V_C(u,t)$ is the prescribed target space for the query. Prediction may
use only interactions strictly preceding $t$ and the textual attributes
available at that time. This formulation treats future target identification
as conditional prediction over evolving interactions rather than binary
classification of independently constructed candidate edges.

\noindent\textbf{Definition 3 (Role-based Inductive Interaction Prediction).}
For a test interaction $(u,v,t)$, let the observed source support be
\[
h(u)=
\left|
\left\{
(u,v',t')\in E_{\mathrm{train}}\cup E_{\mathrm{val}}
\right\}
\right|,
\]
and let $h=h(u)$ when the query is clear. The prediction is
\emph{source-inductive} if $h=0$, meaning that the source has no interaction
in the training or validation graph. When $h>0$, the source is
history-supported, with larger $h$ indicating more available structural and
temporal evidence. The prediction is \emph{target-inductive} if $v$ has not
appeared in its target role during training or validation. In a bipartite
recommendation graph, source- and target-inductive prediction respectively
correspond to new-user and new-item cold-start.

\begin{figure}[t]
\centering
\includegraphics[width=0.95\columnwidth,keepaspectratio]{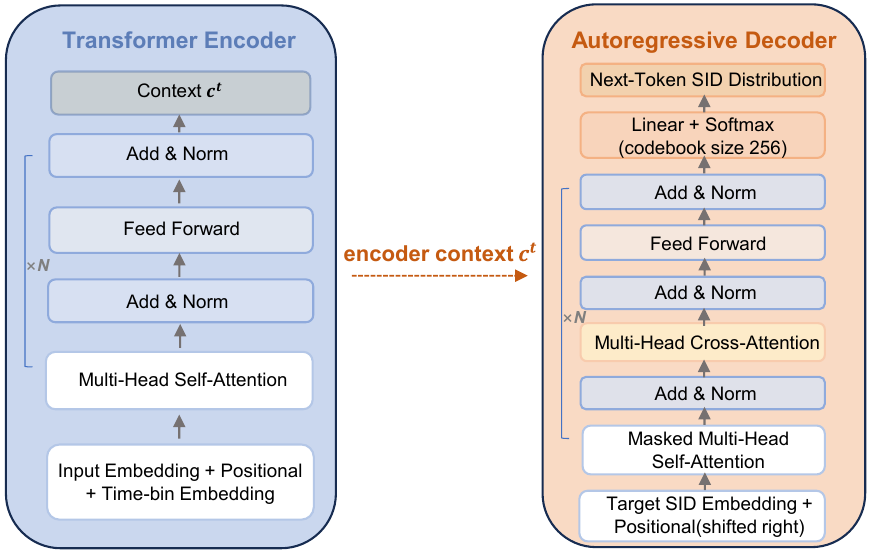}
\caption{Detailed architecture of the Transformer Encoder and Autoregressive Decoder in \method.
The encoder processes the input sequence to produce a context vector $\mathbf{c}^{t}_{u}$; the decoder then generates target SIDs autoregressively, using masked self-attention and cross-attention mechanisms.}
\label{fig:framework_detail}
\end{figure}

% To quantify source-history imbalance, we count each test source's interactions
% in the training and validation periods and divide queries into cold ($h=0$)
% and hot ($h>10$) buckets. Table~\ref{tab:prelim_cold} reports a diagnostic with
% DyGFormer. Its cold-bucket Hits@10 is only 54\%, 17\%, and 2\% of its
% hot-bucket value on Wikirevision, IMDB, and Weibo-Daily, respectively. The
% severity varies by dataset, but the direction is consistent: temporal evidence
% is useful once available, whereas history-only representations are poorly
% supported for cold sources. This motivates a representation that can fall back
% on node text at $h=0$ and incorporate interaction events as $h$ grows.

% \begin{table}[t]
% \centering
% \caption{Motivating source-history diagnostic: filtered Hits@10 of DyGFormer
% for cold ($h=0$) and hot ($h>10$) test sources.}
% \label{tab:prelim_cold}
% \small
% \begin{tabular}{lccc}
% \toprule
% Dataset & cold ($h=0$) & hot ($h>10$) & cold/hot \\
% \midrule
% Wikirevision & 0.195 & 0.360 & 54\% \\
% IMDB         & 0.049 & 0.289 & 17\% \\
% Weibo-Daily  & 0.011 & 0.530 & 2\% \\
% \bottomrule
% \end{tabular}
% \end{table}

\section{Methodology}
\label{sec:method}

\subsection{Overall Framework}
\label{sec:method:framework}

% \method is a generative semantic-ID framework for dynamic link prediction on
% text-attributed graphs. Given a query $(u,t)$, it estimates a target through
% the likelihood of its SID:
% \begin{equation}
%   p(v\mid u,t)\propto p_\theta(\mathbf{s}_v\mid u,t),
%   \label{eq:reformulation}
% \end{equation}
% where $\mathbf{s}_v=(s_v^1,\ldots,s_v^L)$. The framework contains a
% text-to-SID quantizer, an event-level autoregressive generator, and an optional
% inference-time structural calibration. For the original eight datasets, the
% SID contains four base tokens and one disambiguation suffix ($L=5$). The large-scale configuration three base tokens and one suffix ($L=4$); both use a
% base codebook size of $K_c=256$.
\method formulates future-target identification as structured conditional
generation in a discrete semantic space. Given a query $(u,t)$, the
generation component estimates a candidate target through the likelihood of
its SID:
\begin{equation}
p_{\mathrm{gen}}(v\mid u,t)
\propto p_\theta(\mathbf{s}_v\mid \mathbf{s}_u,\mathcal{H}_u^t),
\label{eq:reformulation}
\end{equation}
where $\mathbf{s}_u=(s_u^1,\ldots,s_u^L)$ and
$\mathbf{s}_v=(s_v^1,\ldots,s_v^L)$ are the text-derived SIDs of the
source and candidate target, respectively, and
$\mathcal{H}_u^t$ contains only interactions preceding $t$. Rather than
predicting an independent score from an unconstrained node embedding, the
generator decomposes target identification into a short sequence of
code-token decisions. Semantically related nodes may consequently share part of their generative structure, while the full SID improves candidate-level distinguishability whenever it is collision-free.

The framework integrates three complementary forms of evidence. First,
text-derived SIDs provide a semantic representation that is available even
when a node has no observed interaction. Second, the event sequence supplies
identity-aware temporal evidence through previously interacted targets and
their relative time gaps. Third, structural calibration introduces
graph-specific regularities, including repetition, co-occurrence, and
popularity, without changing the generator or its training objective. These
signals are organized according to their availability rather than assigned to
separate cold-start and history-rich models: the same conditional generator
operates when the history is empty, partially observed, or sufficiently
informative.

Accordingly, \method consists of a text-to-SID quantizer, an event-level
semantic--temporal encoder, an autoregressive SID decoder, and a lightweight
inference-time calibration module. We use a shared base codebook size of $K_c=256$. The base SID length is
$L_{\mathrm{base}}=3$ for Amazon Movies, Wikilife, and Cora and
$L_{\mathrm{base}}=4$ for the remaining six datasets. Appending one finite-vocabulary disambiguation token gives total SID lengths of four and five, respectively.

\subsection{Text-to-SID Quantization}
\label{sec:method:sid}

For each node $v$, a frozen BERT-base encoder~\cite{BERT} produces
\begin{equation}
  \mathbf{e}_v=\operatorname{BERT}(x_v).
\end{equation}
An RQ-VAE~\cite{RQ-VAE22} maps this representation into residual codes. An
MLP encoder first produces
$\mathbf{z}^{(0)}_v=f_{\mathrm{enc}}(\mathbf{e}_v)$. At level
$\ell\in\{1,\ldots,L_{\mathrm{base}}\}$,
\begin{equation}
s_v^\ell=\arg\min_{k\in[K_c]}
\left\|\mathbf{z}_v^{(\ell-1)}-\mathbf{c}_k\right\|_2^2,\quad
\mathbf{z}_v^{(\ell)}
=\mathbf{z}_v^{(\ell-1)}-\mathbf{c}_{s_v^\ell}.
\label{eq:rqvae}
\end{equation}
An MLP decoder reconstructs the input from the selected code vectors, and the
RQ-VAE uses standard reconstruction and commitment losses. We share one
codebook across levels. Because a base SID can be assigned to multiple nodes,
we append a one-token disambiguation suffix that uniquely indexes nodes within
each collision bucket~\cite{pureSID}. The suffix resolves collisions whenever the bucket size is within its finite
vocabulary capacity and otherwise substantially reduces the remaining
ambiguity.

\paragraph{Why use a discrete target representation?}
The SID is not introduced merely as a compressed replacement for a continuous
embedding. It provides a structured prediction space in which semantically
related targets can share early code tokens, while later residual codes refine
their distinction. Consequently, the generator learns token-level regularities
over a codebook of size $K_c$ rather than a separate output class for every
node. This is especially useful when a target has little structural evidence:
its text can still place it near semantically related targets in the shared
code space. At the same time, semantic similarity alone does not guarantee
node identity. The suffix further distinguishes nodes that receive the same base-code tuple,
thereby reducing ambiguity in candidate likelihoods. Extremely large
collision buckets may nevertheless retain residual full-SID collisions under
a finite suffix vocabulary. Our collision and suffix
experiments in Section~\ref{sec:ablation} test these two roles directly. Since
we do not include a continuous-target retrieval control, however, the results
support the usefulness of the SID interface rather than proving that
discretization is universally superior to continuous representations.

\paragraph{Stable and snapshot-conditioned assignments.}
In the stable mode, each node receives one SID from its static text embedding
and retains it across time. For the snapshot-conditioned analysis, the
frozen text embedding is held fixed while the graph-dependent input to the
joint quantizer is recomputed using each temporally preceding graph snapshot.
This produces a time-indexed sequence $\{\mathbf{s}_v^{(r)}\}_r$ under the
same codebook. Consequently, assignment changes in this variant reflect graph
evolution, not changing text. Both modes use the same downstream generator.

\subsection{Event-Level History Encoding}
\label{sec:method:event}

For a query $(u,t)$, we collect the $K$ most recent interactions before $t$:
\begin{equation}
\mathcal{H}_u^t=
\big[(\mathbf{s}_{v_1},\tau_1),\ldots,
(\mathbf{s}_{v_K},\tau_K)\big],
\label{eq:history}
\end{equation}
where $\tau_k=t-t_k$. A logarithmic binning function
$b_k=\phi(\tau_k)\in[B]$ maps time gaps to learnable embeddings. The source
encoder input is
\begin{equation}
\mathbf{X}_u^t=
[\mathbf{s}_u]\Vert
[\mathbf{s}_{v_1},\phi(\tau_1)]\Vert\cdots\Vert
[\mathbf{s}_{v_K},\phi(\tau_K)].
\label{eq:srcseq}
\end{equation}
\paragraph{Identity-aware temporal conditioning.}
The event sequence deliberately retains the identity, order, and relative
time of previous interactions instead of compressing the history into only
its length or a pooled activity vector. Repeated target SIDs expose persistent
preferences, transitions between different SIDs describe changes in
interaction semantics, and time-gap embeddings allow the encoder to
distinguish recent evidence from stale observations. The Transformer can
therefore construct the query representation from both \emph{what} the source
previously interacted with and \emph{when} those interactions occurred.

Sequences shorter than $K$ are padded and masked, while histories longer than
$K$ retain the most recent events. Let $h$ denote the number of source
interactions available before the query. When $h=0$,
$\mathcal{H}_u^t=\varnothing$ and the input reduces to the source SID; no
synthetic neighbor or imputed interaction is introduced. When $h>0$, event
tokens progressively augment the same text-derived representation. This
construction produces a continuous source-side cold-to-hot spectrum without
requiring a separate model or a manually selected switching threshold.

\subsection{Autoregressive Target SID Generation}
\label{sec:method:argen}

An encoder--decoder Transformer maps the source sequence to
\begin{equation}
  \mathbf{c}_u^t=\operatorname{Enc}_\theta(\mathbf{X}_u^t).
  \label{eq:encoder}
\end{equation}
The decoder generates the target SID from left to right:
\begin{equation}
p_\theta(s_v^\ell\mid\mathbf{c}_u^t,s_v^{<\ell})
=\operatorname{softmax}\!\left(
\mathbf{W}_{\mathrm{out}}
\operatorname{Dec}_\theta(\mathbf{c}_u^t,s_v^{<\ell})_\ell
\right).
\label{eq:ardec}
\end{equation}
The candidate likelihood factorizes as
\begin{equation}
p_\theta(\mathbf{s}_v\mid u,t)
=\prod_{\ell=1}^{L}
p_\theta(s_v^\ell\mid\mathbf{c}_u^t,s_v^{<\ell}).
\label{eq:likelihood}
\end{equation}
For training edges $(u,v,t)\in E_{\mathrm{train}}$, we minimize
\begin{equation}
\mathcal{L}_{\mathrm{gen}}(\theta)=-
\sum_{(u,v,t)\in E_{\mathrm{train}}}
\sum_{\ell=1}^{L}
\log p_\theta(s_v^\ell\mid\mathbf{c}_u^t,s_v^{<\ell}).
\label{eq:loss}
\end{equation}
No node-level negative sampling is required during training because each
generation step uses a softmax over the SID codebook.

% \paragraph{Conditioning across the cold-to-hot spectrum.}
% The same likelihood in Eq.~\eqref{eq:likelihood} operates under different amounts of temporal support. When $\mathcal{H}_u^t$ is empty, prediction is conditioned on the source SID, so text provides a nonzero semantic starting point without fabricating historical neighbors. As interactions arrive, each event contributes both the target SID and a relative time-bin embedding; the encoder can therefore distinguish recent preference evidence from distant history. This design does not require separate cold and warm models or a handcrafted switching threshold. The controlled support curve and the randomized-history ablation later test whether additional event context improves ranking and whether the identities of historical targets contribute information beyond history length and timestamps.
\paragraph{Role-based inductive operation.}
Source- and target-inductive prediction stress different parts of the same
generative formulation. For a source-inductive query, the source has no
training or validation interaction, and prediction is therefore initialized
from its text-derived SID without structural history. As interactions
accumulate, their target identities and temporal positions enrich the query
context, but the prediction interface remains unchanged. For a
target-inductive candidate, the source may retain its available history,
whereas the candidate has not appeared previously in its target role.
Provided that its textual attribute is available at query time, the frozen
quantizer can still assign it an SID, whose likelihood is evaluated using the
same decoder as for previously observed targets.

This role-based interpretation separates two distinct generalization
questions. Source-inductive evaluation examines whether the model can form a
useful query representation without source-side interactions, whereas
target-inductive evaluation examines whether text-derived target codes support
the scoring of candidates without target-role interaction evidence. DyGen
handles both cases through the same semantic-ID interface rather than through
task-specific cold-start branches. It does not, however, remove the prescribed
candidate set or claim unrestricted open-world retrieval.

\subsection{Training and Calibrated Inference}
\label{sec:method:sp}
\label{sec:method:algorithm}

We first train the RQ-VAE on frozen text embeddings and then fix the resulting
quantizer and node SIDs. The encoder--decoder Transformer is subsequently
trained on timestamped edges using Eq.~\eqref{eq:loss}, with each source
history containing only events strictly preceding the prediction time.

\begin{algorithm}[t]
\caption{\method Training and Inference Pipeline}
\label{alg:DyGen}
\small
\begin{algorithmic}[1]
\Require DyTAG $G=(V,E,X)$, history length $K$, SID length $L$,
candidate pools $V_C$
\Ensure Trained generator $\theta$ and ranked candidates

\Statex \textbf{// Stage 1: SID quantization}
\For{each node $v\in V$}
  \State $\mathbf{e}_v\gets\operatorname{BERT}(x_v)$
\EndFor
\State train the RQ-VAE on $\{\mathbf{e}_v\}_{v\in V}$
\State assign and freeze $\mathbf{s}_v$ for each $v\in V$

\Statex \textbf{// Stage 2: generator training}
\For{each $(u,v,t)\in E_{\mathrm{train}}$}
  \State collect $\mathcal{H}_u^t$ strictly before $t$
  \State $\mathbf{c}_u^t\gets
  \operatorname{Enc}_\theta(\mathbf{s}_u,\mathcal{H}_u^t)$
  \State update $\theta$ using Eq.~\eqref{eq:loss}
\EndFor

\Statex \textbf{// Stage 3: calibrated inference}
\State compute and freeze structural statistics from
$E_{\mathrm{train}}\cup E_{\mathrm{val}}$
\For{each test query $(u,t)$}
  \State collect $\mathcal{H}_u^t$ strictly before $t$
  \State encode $\mathbf{c}_u^t$ once
  \For{each candidate $v\in V_C$}
    \State obtain $\log p_\theta(\mathbf{s}_v\mid u,t)$ using
    Eq.~\eqref{eq:likelihood}
    \State compute $\operatorname{score}(v\mid u,t)$ using
    Eq.~\eqref{eq:score}
  \EndFor
  \State rank $V_C$ by descending score
\EndFor
\end{algorithmic}
\end{algorithm}

At inference time, the source SID and its recent event history are encoded
once into $\mathbf{c}_u^t$. For each prescribed candidate $v$, its
autoregressive log-likelihood is obtained from
Eq.~\eqref{eq:likelihood}. To complement semantic and temporal compatibility
with recurring graph patterns, we compute three structural signals from the
available training--validation prefix:
$S_{\mathrm{rep}}(u,v)$ measures previous interactions between source $u$ and
candidate $v$;
$S_{\mathrm{co}}(\mathcal{H}_u^t,v)$ measures the co-occurrence of $v$ with
targets in the source history; and
$S_{\mathrm{pop}}(v)$ measures global target popularity.
After normalization, their weights are selected on validation data, and
candidates are ranked by
\begingroup
\setlength{\abovedisplayskip}{4pt}
\setlength{\belowdisplayskip}{4pt}
\setlength{\abovedisplayshortskip}{4pt}
\setlength{\belowdisplayshortskip}{4pt}
\setlength{\jot}{2pt}
\begin{equation}
\begin{aligned}
\operatorname{score}(v\mid u,t)
={}&\log p_\theta(\mathbf{s}_v\mid u,t)
+\lambda_{\mathrm{rep}}S_{\mathrm{rep}}(u,v)\\
&+\lambda_{\mathrm{co}}S_{\mathrm{co}}(\mathcal{H}_u^t,v)
+\lambda_{\mathrm{pop}}S_{\mathrm{pop}}(v).
\end{aligned}
\label{eq:score}
\end{equation}
\endgroup
Setting all three weights to zero recovers generation-only prediction.
Structural calibration changes neither the generator nor its training
objective. Each candidate still requires SID lookup, likelihood aggregation,
calibration, and ranking, but not a separate query-context encoding. Thus,
\method is candidate-efficient rather than candidate-free.
Algorithm~\ref{alg:DyGen} summarizes the complete pipeline.
% \begin{algorithm}[t]
% \caption{\method Training and Inference Pipeline}
% \label{alg:DyGen}
% \small
% \begin{algorithmic}[1]
% \Require DyTAG $G=(V,E,X)$, history length $K$, SID length $L$,
% candidate pools $V_C$
% \Ensure Trained generator $\theta$ and ranked candidates

% \Statex \textbf{// Stage 1: SID quantization}
% \For{each node $v\in V$}
%   \State $\mathbf{e}_v\gets\operatorname{BERT}(x_v)$
% \EndFor
% \State train the RQ-VAE on $\{\mathbf{e}_v\}_{v\in V}$
% \State assign and freeze $\mathbf{s}_v$ for each $v\in V$

% \Statex \textbf{// Stage 2: generator training}
% \For{each $(u,v,t)\in E_{\mathrm{train}}$}
%   \State collect $\mathcal{H}_u^t$ strictly before $t$
%   \State $\mathbf{c}_u^t\gets
%   \operatorname{Enc}_\theta(\mathbf{s}_u,\mathcal{H}_u^t)$
%   \State update $\theta$ using Eq.~\eqref{eq:loss}
% \EndFor

% \Statex \textbf{// Stage 3: calibrated inference}
% \State compute and freeze structural statistics from
% $E_{\mathrm{train}}\cup E_{\mathrm{val}}$
% \For{each test query $(u,t)$}
%   \State collect $\mathcal{H}_u^t$ strictly before $t$
%   \State encode $\mathbf{c}_u^t$ once
%   \For{each candidate $v\in V_C$}
%     \State obtain $\log p_\theta(\mathbf{s}_v\mid u,t)$ using
%     Eq.~\eqref{eq:likelihood}
%     \State compute $\operatorname{score}(v\mid u,t)$ using
%     Eq.~\eqref{eq:score}
%   \EndFor
%   \State rank $V_C$ by descending score
% \EndFor
% \end{algorithmic}
% \end{algorithm}
\section{Experiments}
\label{sec:experiments}
We evaluate \method along several complementary dimensions, including overall performance, inductive and cold-start generalization, component and SID analyses, and computational efficiency. Beyond aggregate results, we further examine how its predictions evolve as interaction evidence accumulates and which model components contribute to this behavior.

\subsection{Experimental Setup}
\label{sec:setup}

\paragraph{Datasets.}
We use eight public DyTAG datasets preprocessed following
GDGB~\cite{GDGB}, with frozen BERT-base text features, and the larger Amazon
Movies graph released by DTGB~\cite{DTGB}. Amazon Movies is derived from the
Movies and TV subset of the Amazon Review Data
collection~\cite{NiLM19Amazon}. It contains 233,459 users, 60,107 items, and
3,217,324 timestamped reviews from May 1996 to July 2014. Chronological
splitting produces 2,252,580 training, 482,480 validation, and 482,264 test
interactions. We use the benchmark-provided static metadata and do not
construct node profiles from held-out reviews. Dataset statistics are
summarized in Table~\ref{tab:datasets}.

\begin{table}[t]
\centering
\caption{Statistics of the nine DyTAG datasets.}
\label{tab:datasets}
\setlength{\tabcolsep}{3.2pt}
\begin{tabular}{lrrcc}
\toprule
Dataset & \#Nodes & \#Edges & $M/N$ & Domain \\
\midrule
Wikirevision & 4{,}243   & 104{,}807    & 24.7 & Wiki edits \\
Amazon Movies & 293{,}566 & 3{,}217{,}324 & 11.0 & Movie reviews \\
IMDB         & 24{,}310  & 211{,}497    & 8.7  & Movie credits \\
Dianping     & 11{,}837  & 66{,}283     & 5.6  & Reviews \\
Weibo-Daily  & 19{,}768  & 71{,}165     & 3.6  & Social reposts \\
Weibo-Tech   & 25{,}613  & 92{,}209     & 3.6  & Social reposts \\
Wikilife     & 22{,}403  & 67{,}209     & 3.0  & Wiki edits \\
Sephora      & 26{,}108  & 67{,}881     & 2.6  & Reviews \\
Cora         & 49{,}217  & 108{,}277    & 2.2  & Citations \\
\bottomrule
\end{tabular}
\end{table}

\paragraph{Evaluation protocol.}
For each of the original eight datasets, we sample 2,000 test queries and
rank the ground-truth target against 4,999 uniformly sampled negatives.
We report filtered Hits@$K$ after removing candidates that form known
training or validation interactions with the query source. On Amazon, every
method uses the same fixed manifest of 2,000 queries and 5,000 target-only
candidates, with expected random tie-breaking. Unless stated otherwise,
model inputs use only interactions preceding the query, and structural
statistics are computed from the training--validation prefix and frozen
before testing. The held-out review text and rating are not used as model
inputs. The main evaluation is sampled ranking rather than full-catalog
retrieval.

\paragraph{Baselines and implementation.}
We compare with EdgeBank~\cite{EdgeBank}, GraphMixer~\cite{GraphMixer},
TGN~\cite{TGN}, TPNet~\cite{TPNet}, DyGFormer~\cite{DyGLib}, and the
generative baseline TIGER~\cite{TIGER}, using their recommended
configurations. TGN exhausts the available 24\,GB GPU memory on Amazon and is
reported as OOM rather than used for an accuracy comparison.

We use a shared base codebook size of $K_c=256$. The base SID length is
$L_{\mathrm{base}}=3$ for Amazon Movies, Wikilife, and Cora and
$L_{\mathrm{base}}=4$ for the remaining six datasets. Appending one
disambiguation token gives total SID lengths of four and five, respectively. The generator
comprises a four-layer Transformer encoder and a four-layer Transformer
decoder, with $d_{\mathrm{model}}=256$, eight attention heads, an FFN hidden
dimension of 1024, and a maximum history length of 32. On Amazon, the
generator is trained for 60 epochs with a history-masking ratio of 0.2, a
corresponding loss weight of 1.5, and a temporal regularization weight of
0.1. The validation-selected calibration weights are
$\lambda_{\mathrm{rep}}=8.0$, $\lambda_{\mathrm{co}}=0.5$, and
$\lambda_{\mathrm{pop}}=0.5$. All experiments use one NVIDIA RTX 3090 GPU.

\paragraph{Statistical analysis.}
For the original eight datasets, Hits@10 over 2,000 queries is treated as a
binary query-level outcome. We report Wilson confidence intervals and
Wilson-MOVER intervals for differences, and compare \method with TIGER using
two-sided two-proportion score tests with Holm correction across datasets.
Because the historical exports do not preserve identical candidate arrays
across methods, these comparisons are treated as independent-sample
approximations. Amazon comparisons use 10,000 paired query-bootstrap
resamples over identical queries and candidate sets, with Holm correction
within each reported comparison family. These intervals quantify test-query
uncertainty for fixed trained models rather than variation across training
seeds.
\begin{figure}[t]
\centering
\includegraphics[width=0.95\linewidth]{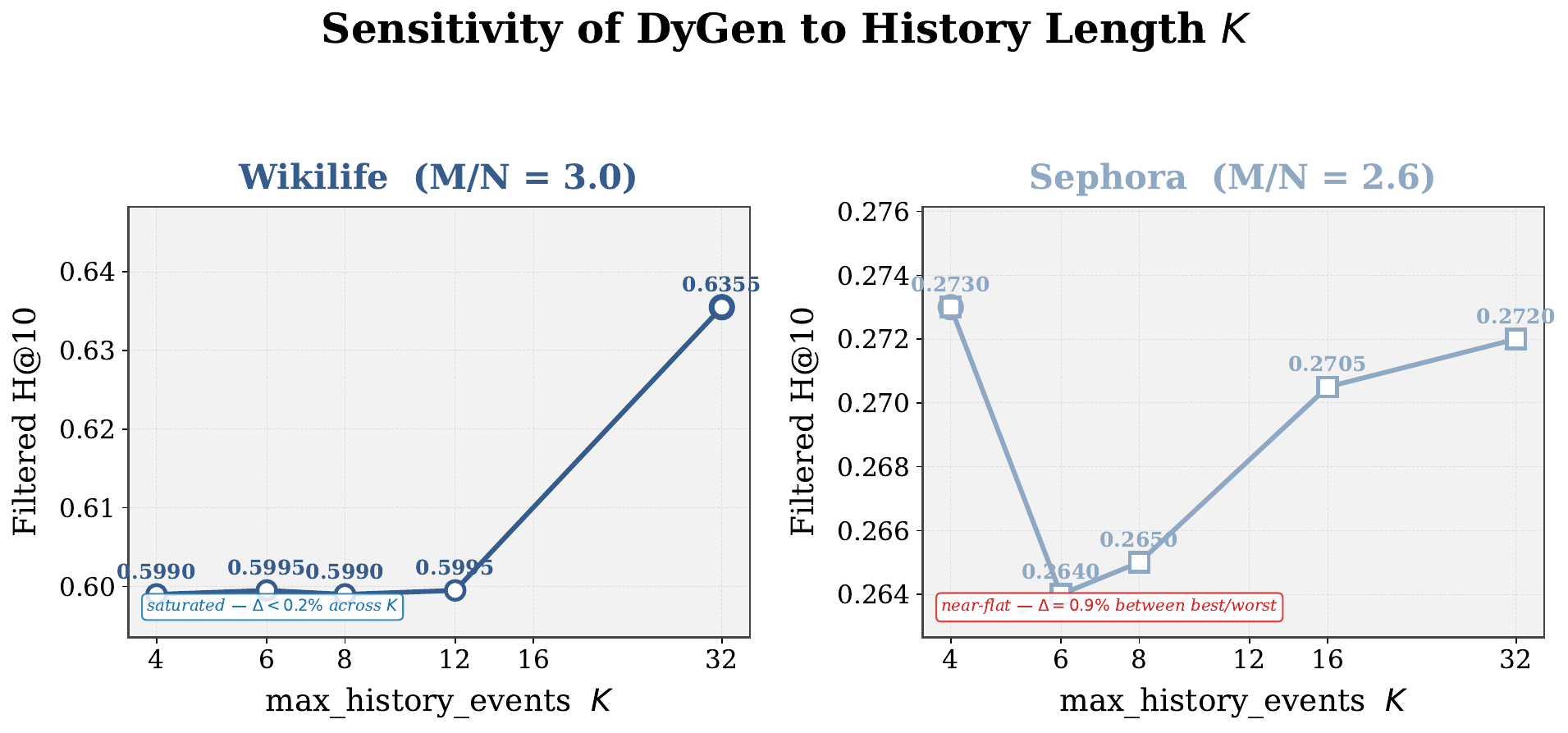}
\caption{Sensitivity of \method to history length $K$ on Wikilife
($M/N=3.0$) and Sephora ($M/N=2.6$). Wikilife benefits from the longest tested
history, whereas Sephora is comparatively insensitive to $K$.}
\label{fig:k_sensitivity}
\end{figure}

\subsection{Overall Results}
\label{sec:main_results}

\begin{table*}[t]
\centering
\caption{Filtered Hits@10 on nine DyTAG datasets, ordered by $M/N$.
\textsuperscript{\(\dagger\)} marks a Holm-corrected significant gain over
TIGER.}
\label{tab:main}
\small
\setlength{\tabcolsep}{4.1pt}
\begin{tabular}{lrccccccc}
\toprule
Dataset & $M/N$ & EdgeBank & GraphMixer & TGN & DyGFormer & TPNet & TIGER & \method \\
\midrule
Wikirevision & 24.7 & \underline{0.3200} & 0.2290 & OOM
% & 0.3060 & 0.2110 & 0.2990 & \textbf{0.5170} \\
& 0.3060 & 0.2110 & 0.2990
& \textbf{0.5170}\textsuperscript{\(\dagger\)} \\
Amazon Movies & 11.0 & 0.0659 & 0.0625 & OOM
& 0.1233 & \textbf{0.1460} & 0.0037 & \underline{0.1455} \\
IMDB & 8.7 & 0.0460 & 0.0740 & OOM
% & \textbf{0.2055} & 0.0565 & 0.0585 & \underline{0.1425} \\
& \textbf{0.2055} & 0.0565 & 0.0585
& \underline{0.1425}\textsuperscript{\(\dagger\)} \\
Dianping & 5.6 & 0.0015 & \textbf{0.1935} & OOM
% & 0.1130 & 0.0380 & 0.0955 & \underline{0.1770} \\
& 0.1130 & 0.0380 & 0.0955
& \underline{0.1770}\textsuperscript{\(\dagger\)} \\
Weibo-Daily & 3.6 & \underline{0.4190} & 0.0825 & OOM
& 0.4105 & 0.3560 & 0.4095 & \textbf{0.4430} \\
Weibo-Tech & 3.6 & \textbf{0.4890} & 0.0695 & 0.1680
% & 0.4350 & 0.2080 & 0.2965 & \underline{0.4730} \\
& 0.4350 & 0.2080 & 0.2965
& \underline{0.4730}\textsuperscript{\(\dagger\)} \\
Wikilife & 3.0 & 0.1150 & 0.4315 & OOM
& \underline{0.5260} & 0.2735 & \textbf{0.6355} & \textbf{0.6355} \\
Sephora & 2.6 & 0.0020 & 0.0355 & OOM
& 0.2370 & 0.0220 & \textbf{0.2930} & \underline{0.2730} \\
% Cora & 2.2 & 0.0015 & \textbf{0.1585} & \underline{0.1470}
% & 0.0475 & 0.0150 & 0.0800 & 0.1080 \\
Cora & 2.2 & 0.0015 & \textbf{0.1585} & \underline{0.1470}
& 0.0475 & 0.0150 & 0.0800
& 0.1080\textsuperscript{\(\dagger\)} \\
\bottomrule
\end{tabular}
\end{table*}

\begin{table}[t]
\centering
\caption{Amazon Movies performance on 2,000 queries with 5,000 candidates.
\textsuperscript{*} marks a Holm-corrected significant gain over the strongest
non-\method result for the corresponding metric.}
\label{tab:amazon-main}
\small
\begin{tabular}{lccccc}
\toprule
Method & Hits@1 & Hits@5 & Hits@10 & Hits@20 & MRR \\
\midrule
EdgeBank & \underline{0.0642} & 0.0649 & 0.0659 & 0.0677 & 0.0657 \\
GraphMixer & 0.0125 & 0.0380 & 0.0625 & 0.1040 & 0.0326 \\
TGN & OOM & OOM & OOM & OOM & OOM \\
DyGFormer & 0.0315 & 0.0820 & 0.1233 & 0.1770 & 0.0645 \\
TPNet & 0.0435 & \underline{0.1030} & \textbf{0.1460}
& \textbf{0.2010} & \underline{0.0801} \\
TIGER & 0.0003 & 0.0017 & 0.0037 & 0.0072 & 0.0026 \\
\midrule
\method & \textbf{0.0830}\textsuperscript{*}
& \textbf{0.1200}\textsuperscript{*}
& \underline{0.1455} & \underline{0.1805}
& \textbf{0.1064}\textsuperscript{*} \\
\bottomrule
\end{tabular}
\end{table}

Table~\ref{tab:main} shows that \method ranks first or ties for first on
three of the original eight datasets and is among the top two on seven.
Including Amazon, it is among the top two on eight of nine datasets. The
method is therefore broadly competitive, rather than uniformly best. Compared
with TIGER, the largest gains occur on denser graphs, where event histories
provide useful context; on sparse graphs, \method approaches the behavior of
the static-SID baseline.

\paragraph{Benefits of event-level conditioning.}
Beyond point estimates, \method significantly outperforms the static-SID
generative baseline TIGER on five of the eight original datasets after Holm
correction: Wikirevision, IMDB, Dianping, Weibo-Tech, and Cora. The differences
on Weibo-Daily, Wikilife, and Sephora are not statistically significant. In
particular, the positive difference on Weibo-Daily does not remain significant
after correction, Wikilife is an exact point-estimate tie, and the negative
difference on Sephora is also nonsignificant. Thus, none of the original
datasets provides evidence that \method is significantly worse than TIGER.
Together with the \textbf{RandHist} intervention in
Section~\ref{sec:ablation}, these results are consistent with the value of
augmenting static semantic-ID generation with identity-aware temporal
context, although the benefit is not uniform across graph regimes.

\paragraph{Complementary baseline behavior.}
The table also exposes distinct strengths across model families. DyGFormer is
strongest on IMDB, GraphMixer on Dianping and Cora, and the memory-based
EdgeBank on Weibo-Tech. These exceptions are important: they show that no
single inductive bias dominates every density and recurrence regime. \method
is most useful when semantic identifiers and event-level history provide
complementary evidence, while simpler structural or memory-based methods can
remain preferable on particular graphs.

On Amazon, Table~\ref{tab:amazon-main} shows that \method performs best on
Hits@1, Hits@5, and MRR. Its Hits@10 is nearly identical to TPNet
($0.1455$ versus $0.1460$), while TPNet has the higher Hits@20 point estimate.
Against TPNet, the improvements on Hits@1 ($\Delta=0.0395$,
$p_{\mathrm{adj}}<0.001$), Hits@5 ($\Delta=0.0170$,
$p_{\mathrm{adj}}=0.0384$), and MRR ($\Delta=0.0263$,
$p_{\mathrm{adj}}<0.001$) are significant. Hits@10
($p_{\mathrm{adj}}=0.9871$) and Hits@20
($p_{\mathrm{adj}}=0.0912$) are not.

\paragraph{Significance beyond the strongest baseline.}
The paired analysis gives a consistent but metric-dependent picture. Against
DyGFormer, \method is significantly better on Hits@1, Hits@5, Hits@10, and
MRR (adjusted $p\leq0.0260$), whereas Hits@20 is not significant
($p_{\mathrm{adj}}=0.7411$). Against GraphMixer, EdgeBank, and TIGER, all
five improvements are significant after Holm correction. Together with the
TPNet comparison, these results locate the main Amazon advantage at early-rank
precision and reciprocal rank rather than claiming a uniform gain at every
cutoff. They also provide evidence at a graph roughly six times larger in
nodes and fifteen times larger in interactions than the largest original
benchmark along the corresponding dimension.

% WSDM PAGE-LIMIT CUT: the supplied source contained a duplicate
% Analysis of Main Experimental Results subsection before Table~\ref{tab:main}.
% Its density-regime discussion is condensed into the two paragraphs above.

\subsection{Inductive and Recommendation Cold-Start Evaluation}
\label{sec:coldstart}

% \begin{table}[t]
% \centering
% \caption{Source-side inductive link prediction: filtered Hits@10 for test
% queries whose source has no training or validation interactions ($h=0$).
% This setting evaluates generalization to structurally unobserved sources
% and does not imply target-side novelty. Best is in \textbf{bold} and
% second-best is \underline{underlined}. ``--'' indicates the method has no
% cold-bucket queries in its evaluation sample.}
% \label{tab:cold}
% \small
% \setlength{\tabcolsep}{3.1pt}
% \begin{tabular}{lrcccc}
% \toprule
% Dataset & $M/N$ & GraphMixer & DyGFormer & TIGER & \method \\
% \midrule
% Wikirevision & 24.7
% & 0.1913
% & 0.1946
% & \underline{0.2104}
% & \textbf{0.2452} \\

% Amazon Movies & 11.0
% & \underline{0.0412}
% & 0.0262
% & 0.0027
% & \textbf{0.0562} \\

% IMDB & 8.7
% & \textbf{0.0597}
% & \underline{0.0494}
% & 0.0267
% & 0.0411 \\

% Dianping & 5.6
% & \textbf{0.2563}
% & 0.1744
% & 0.1455
% & \underline{0.2328} \\

% Weibo-Daily & 3.6
% & 0.0054
% & 0.0108
% & \underline{0.0139}
% & \textbf{0.0251} \\

% Weibo-Tech & 3.6
% & 0.0325
% & 0.0325
% & \underline{0.0339}
% & \textbf{0.0508} \\

% Wikilife & 3.0
% & 0.3694
% & \underline{0.4192}
% & --
% & \textbf{0.5500} \\

% Sephora & 2.6
% & 0.0376
% & \underline{0.3906}
% & \textbf{0.4017}
% & 0.3806 \\
% \bottomrule
% \end{tabular}
% \end{table}
\begin{figure}[t]
    \centering
    \includegraphics[width=0.47\textwidth]
    {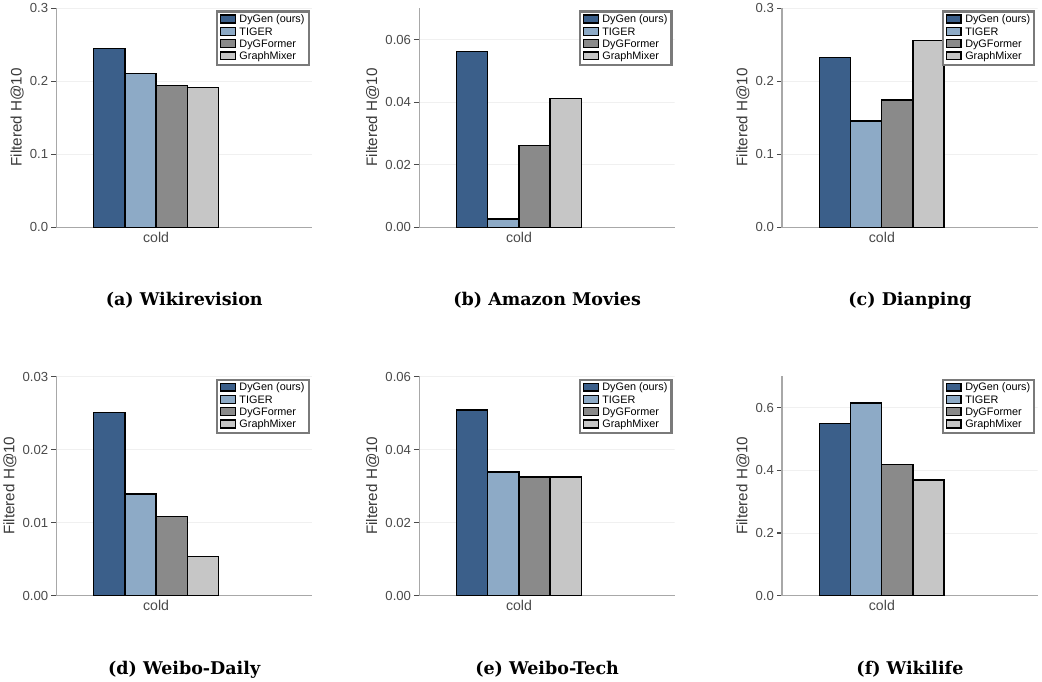}
    \caption{Source-side inductive performance on six representative
    datasets. Bars report filtered Hits@10 for test queries whose sources
    have no interactions during training or validation ($h=0$). $M/N$ is reported in each panel title.}
    \label{fig:source-inductive}
\end{figure}
\begin{table*}[t]
\centering
\caption{Cold-start evaluation on Amazon Movies.
(a) The same query for each of 2,000 new users is evaluated with nested
support sets containing 0, 1, 3, or 5 strictly earlier interactions;
each query has 5,000 candidates.
(b) For 186 natural new-item first-arrival queries, every method ranks all
200 items unseen as targets during training and validation.}
\label{tab:amazon-coldstart}
\small
\setlength{\tabcolsep}{4.2pt}
\begin{tabular}{c cc @{\hspace{1.5em}} l ccccc}
\toprule
\multicolumn{3}{c}{(a) New-user support curve}
& \multicolumn{6}{c}{(b) Natural new-item ranking} \\
\cmidrule(lr){1-3}\cmidrule(lr){4-9}
Support & Hits@10 & MRR
& Method & Hits@1 & Hits@5 & Hits@10 & Hits@20 & MRR \\
\midrule
0 & 0.0380 & 0.0199
& \method & \textbf{0.0108} & \textbf{0.0753}
& \textbf{0.1075} & \textbf{0.1882} & \textbf{0.0546} \\
1 & 0.0875 & 0.0424
& GraphMixer & \textbf{0.0108} & 0.0430
& 0.0645 & 0.1452 & 0.0369 \\
3 & 0.1080 & 0.0519
& DyGFormer & 0.0054 & 0.0269
& 0.0430 & 0.1075 & 0.0307 \\
5 & \textbf{0.1115} & \textbf{0.0573}
& TPNet & 0.0000 & 0.0269
& 0.0323 & 0.1022 & 0.0262 \\
\multicolumn{3}{c}{}
& Random & 0.0050 & 0.0250 & 0.0500 & 0.1000 & 0.0294 \\
\bottomrule
\end{tabular}
\end{table*}
Figure~\ref{fig:source-inductive} presents source-side inductive results
on six representative datasets spanning different graph densities.
\method remains competitive across these settings and obtains strong
point estimates even when the source has no observed interaction during
training or validation. The results suggest that text-derived SIDs provide
a useful history-independent signal for structurally unobserved sources,
although the relative advantage varies with the interaction patterns of different datasets. This evaluation concerns source-side structural
generalization and does not imply target-side novelty.
The Amazon subset provides a particularly controlled source-inductive
setting: all 267 queries have $h=0$, while every ground-truth target has
already appeared during training or validation. Consequently, performance
differences cannot be attributed to target-side novelty. \method obtains the
highest Hits@10 point estimate of 0.0562, compared with 0.0412 for
GraphMixer, 0.0262 for DyGFormer, and 0.0027 for TIGER. Under 10,000 paired
query-bootstrap resamples, the advantage over TIGER is significant
($\Delta=0.0535$, 95\% CI $[0.0284,0.0824]$,
$p_{\mathrm{adj}}=0.0006$). The differences from GraphMixer
($\Delta=0.0150$, 95\% CI $[-0.0187,0.0487]$,
$p_{\mathrm{adj}}=0.4538$) and DyGFormer
($\Delta=0.0300$, 95\% CI $[0.0000,0.0599]$,
$p_{\mathrm{adj}}=0.1344$) are not statistically significant.

Table~\ref{tab:amazon-coldstart}(a) evaluates the same query while gradually
revealing support. The first interaction gives the largest improvement:
Hits@10 increases from 0.0380 to 0.0875, with paired difference 0.0495 and
95\% CI $[0.0380,0.0615]$; MRR simultaneously improves by 0.0224 with
CI $[0.0161,0.0290]$. Moving from one to three interactions yields a further
Hits@10 gain of 0.0205 with CI $[0.0090,0.0320]$. The five-versus-three
Hits@10 difference is only 0.0035 and its interval crosses zero, indicating
saturation at this cutoff, although MRR retains a small positive increment.
Thus, text supplies the zero-support starting point, the first observed event
contributes the largest refinement, and later events provide diminishing
returns. In panel (b), \method has the strongest point estimates on
Hits@5, Hits@10, Hits@20, and MRR and ties GraphMixer on Hits@1. It
significantly outperforms TPNet on Hits@10
($p_{\mathrm{adj}}=0.0104$) and MRR ($p_{\mathrm{adj}}<0.001$), but not
GraphMixer. It also significantly exceeds the random ranker on Hits@5,
Hits@10, Hits@20, and MRR, while the Hits@10 comparison with DyGFormer is
borderline but nonsignificant after correction ($p_{\mathrm{adj}}=0.0512$).
This is full ranking within the natural cold-item universe, not the complete
warm-and-cold catalog.
\vspace{-5pt}
\paragraph{Mixed-catalog calibration diagnostic.}
We additionally stress-test the same 186 new-item targets using the earlier
5,000-candidate construction that mixes warm and cold items. In this manifest,
99.64\% of negatives have already appeared during training or validation, and
an average of 4,416 warm candidates are ranked ahead of the true new item.
The resulting Hits@10 is 0 and MRR is 0.00024. This negative result should not
be interpreted as an inability to assign a semantic SID to a new target: the
cold-only experiment shows meaningful discrimination among unseen items.
Instead, it exposes a score-calibration gap between targets with abundant
interaction evidence and first-arrival targets. We therefore report the
200-item cold-universe ranking as the controlled new-item result and treat
mixed-catalog calibration as an unresolved deployment problem rather than
claiming arbitrary open-catalog retrieval.

\begin{figure}[t]
    \centering
    \subfigure[Wikirevision]{
        \includegraphics[width=0.47\linewidth]{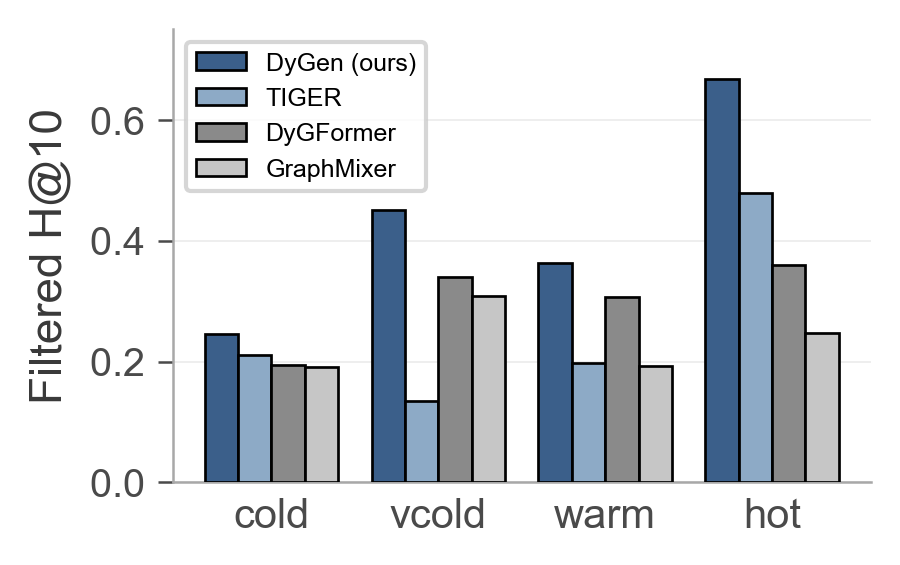}
    }
    \subfigure[Weibo-Daily]{
        \includegraphics[width=0.47\linewidth]{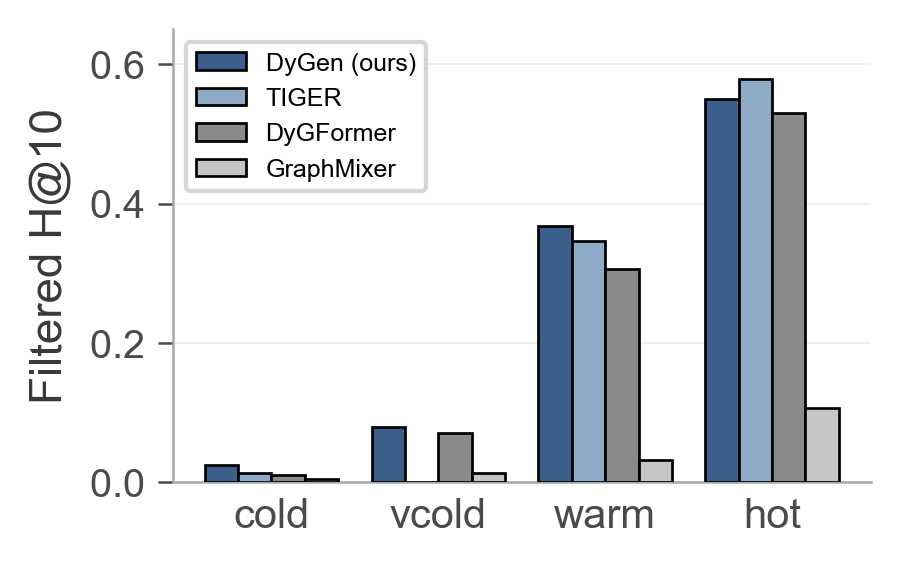}
    }
    \subfigure[Weibo-Tech]{
        \includegraphics[width=0.47\linewidth]{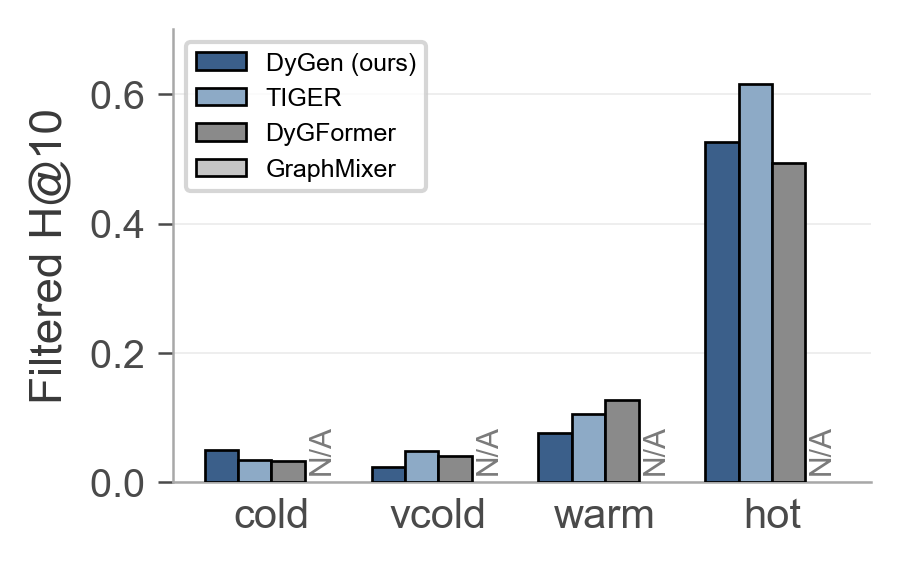}
    }
    \subfigure[Wikilife]{
        \includegraphics[width=0.47\linewidth]{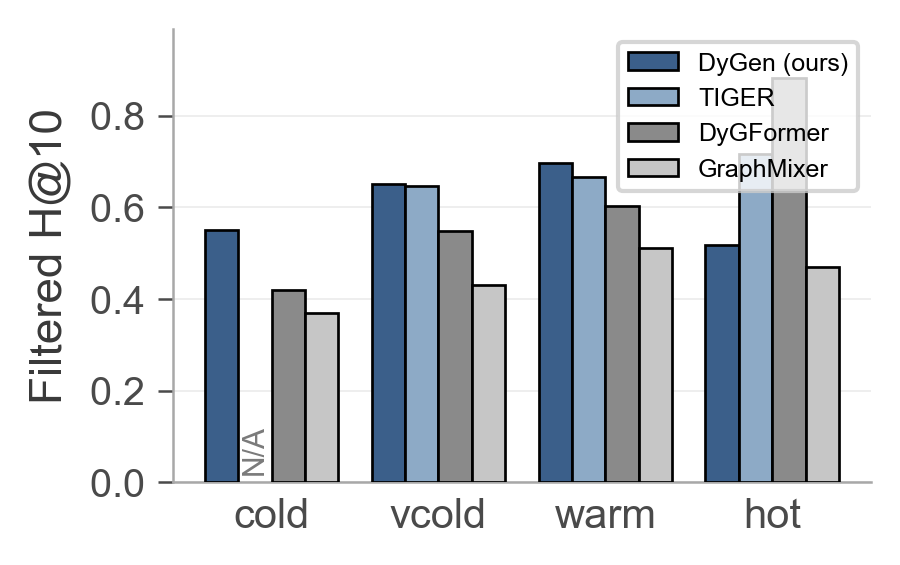}
    }
    \caption{Filtered Hits@10 across source-history buckets on four
    representative datasets. \method remains competitive across the
    cold-to-hot spectrum and is strongest in the $h=0$ source-history bucket
    on these datasets.}
    \label{fig:cold_start_four_datasets}
\end{figure}

For a finer-grained view, Figure~\ref{fig:cold_start_four_datasets} divides
test sources into cold ($h=0$), very cold ($1\leq h\leq2$), warm
($3\leq h\leq10$), and hot ($h>10$) buckets. Across the four representative
datasets, \method remains competitive throughout the cold-to-hot spectrum and
is strongest in the $h=0$ bucket. The relative advantage is most visible on
graphs where the hot bucket is well supported, suggesting that text-derived
SIDs supply the missing signal for cold sources while the same model can still
exploit interaction history for warmer ones. Cora is an important boundary
case because all test sources are cold: \method reaches 0.1080 Hits@10, below
GraphMixer's 0.1585 but above DyGFormer's 0.0475, showing useful text-based
generalization without claiming uniform dominance.

\subsection{Component and SID Analyses}
\label{sec:ablation}
We compare the full model with two sequential interventions.
\textbf{w/o Cal.} disables structural calibration while retaining the original
event histories. Starting from this generation-only model,
\textbf{RandHist} additionally randomizes historical target identities while
preserving history length, event order, positions, and timestamps.
Accordingly, $\Delta_F$ measures the relative change from the full model,
whereas $\Delta_G$ measures the relative change from the generation-only
model (\textbf{w/o Cal.}).
\FloatBarrier
\begin{table}[!ht]
\centering
\caption{Filtered Hits@10 component analysis across nine datasets.
Parentheses report rounded relative changes.}
\label{tab:ablation}
\small
\setlength{\tabcolsep}{4.0pt}
\begin{tabular}{lrcccc}
\toprule
Dataset & $M/N$ & Full \method
& w/o Cal. ($\Delta_F$)
& RandHist ($\Delta_G$) \\
\midrule
Wikirevision & 24.7 & \textbf{0.5170} & 0.4380$_{\downarrow15\%}$ & 0.1650$_{\downarrow62\%}$ \\
Amazon Movies & 11.0 & \textbf{0.1455} & 0.1085$_{\downarrow25\%}$ & 0.0695$_{\downarrow36\%}$ \\
IMDB & 8.7 & \textbf{0.1425} & 0.1055$_{\downarrow26\%}$ & 0.0215$_{\downarrow80\%}$ \\
Dianping & 5.6 & 0.1770 & \textbf{0.1785}$_{\uparrow1\%}$ & 0.0875$_{\downarrow51\%}$ \\
Weibo-Daily & 3.6 & \textbf{0.4430} & 0.2855$_{\downarrow36\%}$ & 0.0405$_{\downarrow86\%}$ \\
Weibo-Tech & 3.6 & \textbf{0.4730} & 0.2485$_{\downarrow47\%}$ & 0.0165$_{\downarrow93\%}$ \\
Wikilife & 3.0 & \textbf{0.6355} & 0.6000$_{\downarrow6\%}$ & 0.5335$_{\downarrow11\%}$ \\
Sephora & 2.6 & 0.2730 & \textbf{0.2725}$_{\downarrow0\%}$ & 0.2675$_{\downarrow2\%}$ \\
Cora & 2.2 & \textbf{0.1080} & 0.1070$_{\downarrow1\%}$ & 0.1070$_{\downarrow1\%}$ \\
\bottomrule
\end{tabular}
\end{table}

Table~\ref{tab:ablation} follows a sequential component intervention. Starting from the full model, \textbf{w/o Cal.} removes structural calibration while retaining the original event histories, thereby recovering generation-only prediction. The full model has a higher Hits@10 point estimate than \textbf{w/o Cal.} on seven of the nine datasets. The two exceptions are Dianping, where generation only is higher by 0.0015, and Sephora, where the difference is only 0.0005. Structural calibration is therefore beneficial across most datasets, but its effect is marginal or slightly negative in particular graph regimes rather than uniformly positive.

RandHist provides a stricter diagnostic of the information carried by
historical target identities. It preserves the number, order, positions, and timestamps of historical events, but replaces their target identities at random and uses no structural calibration. Relative to generation only, RandHist reduces Hits@10 on eight of the nine datasets, with relative decreases ranging from 2\% on Sephora to 93\% on Weibo-Tech; Cora is essentially unchanged. Because the amount and temporal placement of history remain fixed, these results show that \method generally exploits the identities of past interaction targets rather than merely history length or temporal positions.

On Amazon Movies, RandHist reduces Hits@10 from 0.1085 to 0.0695 and MRR from 0.0635 to 0.0387. Both reductions are significant under paired query bootstrap with Holm correction ($p_{\mathrm{adj}}<0.001$).

% Table~\ref{tab:amazon-structural} isolates the three structural signals on
% Amazon. Repetition is the dominant calibration term, co-occurrence provides a
% smaller complementary gain, and popularity alone leaves Hits@10 unchanged.
% The full improvement is therefore not explained by globally frequent targets.

% \begin{table}[t]
% \centering
% \caption{Structural calibration on Amazon under the same 2,000-query,
% 5,000-candidate manifest.}
% \label{tab:amazon-structural}
% \small
% \begin{tabular}{lcc}
% \toprule
% Variant & Hits@10 & $\Delta$ vs. generation \\
% \midrule
% Generation only & 0.1085 & -- \\
% $+$ repetition & 0.1435 & $+0.0350$ \\
% $+$ co-occurrence & 0.1100 & $+0.0015$ \\
% $+$ popularity & 0.1085 & $+0.0000$ \\
% $+$ all signals & \textbf{0.1455} & $+0.0370$ \\
% \bottomrule
% \end{tabular}
% \end{table}

\begin{table}[t]
\centering
\caption{SID design analysis on Amazon. All rows use pure generation on the
same 2,000-query, 5,000-candidate manifest. $L_{\mathrm{base}}$ excludes the
suffix. $\dagger$ marks a diagnostic best-finite checkpoint because later
no-suffix training became non-finite.}
\label{tab:sid-analysis}
\small
\setlength{\tabcolsep}{2.4pt}
\begin{tabular}{lcccccc}
\toprule
Configuration & $K_c$ & $L_{\mathrm{base}}$ & Suffix & $L$ & H@10 & MRR \\
\midrule
Total length 3 & 256 & 2 & Yes & 3 & 0.0995 & 0.0528 \\
Snapshot-cond. & 256 & 3 & Yes & 4 & \textbf{0.1085} & \textbf{0.0635} \\
Total length 5 & 256 & 4 & Yes & 5 & 0.0970 & 0.0564 \\
Larger codebook & 512 & 3 & Yes & 4 & 0.1015 & 0.0559 \\
No suffix\textsuperscript{$\dagger$} & 256 & 3 & No & 3 & 0.0890 & 0.0501 \\
Static SID & 256 & 3 & Yes & 4 & 0.1060 & 0.0608 \\
\bottomrule
\end{tabular}
\end{table}
% \begin{table}[t]
% \centering
% \caption{Collision and temporal stability of Amazon target SIDs. Temporal
% rates are computed over ten training snapshots.}
% \label{tab:sid-mechanism}
% \small
% \setlength{\tabcolsep}{4pt}
% \begin{tabular}{lc}
% \toprule
% Diagnostic & Value \\
% \midrule
% Unique base SIDs & 56,550 / 60,107 (94.08\%) \\
% Base collision buckets & 2,930 \\
% Items in collision buckets & 10.79\% \\
% Maximum collision-bucket size & 14 \\
% Targets using nonzero suffix & 5.92\% \\
% Unique complete SIDs & 60,107 / 60,107 (100\%) \\
% \midrule
% Mean adjacent base-SID change & 3.94\% \\
% Mean adjacent complete-SID change & 4.38\% \\
% Targets changing at least once & 34.04\% \\
% Targets changing at least twice & 4.77\% \\
% \bottomrule
% \end{tabular}
% \end{table}
\paragraph{Cross-dataset SID audit.}
We audit the SIDs over the valid candidate universe of each frozen ranking
protocol: all nodes for the original eight datasets and target items for the
bipartite Amazon Movies dataset. Let $U_{\mathrm{base}}$ and
$U_{\mathrm{full}}$ denote the numbers of distinct base and full SID
sequences divided by the number of candidate nodes, respectively.
$C_{\mathrm{base}}$ is the percentage of candidate nodes belonging to
non-singleton base-SID buckets, $B_{\max}$ is the largest such bucket, and
$R_{\mathrm{adj}}$ is the mean percentage of candidate nodes whose base SID
changes between adjacent training snapshots. $T$ is the number of audited training snapshots and
$L_{\mathrm{base}}$ is the number of base SID tokens before appending the
disambiguation suffix. We report base-SID change because
it is available uniformly across all datasets; complete-SID snapshot changes
are unavailable for IMDB.
\begin{table}[t]
\centering
\caption{Cross-dataset SID collision and temporal-stability audit.}
\label{tab:sid-mechanism}
\small
\setlength{\tabcolsep}{4.2pt}
\begin{tabular}{@{}lrrrrrrr@{}}
\toprule
Dataset
& $T$
& $L_{\mathrm{base}}$
& \shortstack{$U_{\mathrm{base}}$\\(\%)}
& \shortstack{$C_{\mathrm{base}}$\\(\%)}
& $B_{\max}$
& \shortstack{$U_{\mathrm{full}}$\\(\%)}
& \shortstack{$R_{\mathrm{adj}}$\\(\%)} \\
\midrule
Wikirevision & 10 & 4 & 82.75 & 20.15 & 85    & 100.00 & 0.00 \\
Amazon Movies & 10 & 3 & 94.08 & 10.79 & 14  & 100.00 & 3.94 \\
IMDB          & 6  & 4 & 98.01 & 3.61  & 10    & 100.00 & 7.35 \\
Dianping      & 10 & 4 & 88.78 & 14.40 & 7,874 & 95.17  & 0.00 \\
Weibo-Daily   & 10 & 4 & 98.72 & 2.42  & 14    & 100.00 & 0.00 \\
Weibo-Tech    & 10 & 4 & 98.94 & 2.03  & 10    & 100.00 & 0.00 \\
Wikilife      & 6  & 3 & 84.36 & 25.03 & 321   & 99.98  & 3.94 \\
Sephora       & 8  & 4 & 59.18 & 51.83 & 1,197 & 97.15  & 0.00 \\
Cora          & 5  & 3 & 99.06 & 1.84  & 7     & 100.00 & 1.48 \\
\bottomrule
\end{tabular}
\end{table}
% \paragraph{Collision and suffix.}
% Table~\ref{tab:sid-mechanism} shows that the base code already distinguishes
% 94.08\% of Amazon targets, but the remaining collisions are operationally
% important because autoregressive likelihood alone cannot distinguish two
% nodes with the same complete output sequence. A nonzero suffix is required by
% only 5.92\% of targets and makes every complete SID unique. Removing it decreases
% Hits@10 from 0.1085 to 0.0890 and MRR from 0.0635 to 0.0501; all five metric
% degradations are significant after Holm correction. In particular, the
% Hits@10 difference is 0.0195 with 95\% CI $[0.0100,0.0295]$, and the MRR
% difference is 0.0133 with CI $[0.0075,0.0193]$. The suffix therefore
% improves both identifiability and ranking, although the no-suffix comparison
% uses the diagnostic checkpoint policy noted in Table~\ref{tab:sid-analysis}.
\vspace{-5pt}
\paragraph{Collision and suffix.}
Table~\ref{tab:sid-mechanism} shows substantial variation in base-SID
collisions across datasets. The percentage of candidate nodes in
non-singleton base-SID buckets ranges from 1.84\% on Cora to 51.83\% on
Sephora. Appending the disambiguation suffix consistently increases SID
distinctness on all nine datasets: $U_{\mathrm{full}}$ ranges from 95.17\%
to 100\%, with six datasets reaching 100\%. These quantities measure the
ratio of distinct SID sequences to candidate nodes, rather than the percentage
of nodes possessing globally unique identifiers.

The largest collision buckets occur on Dianping, Sephora, and Wikilife,
where $B_{\max}$ is 7,874, 1,197, and 321, respectively. Their full-SID
distinctness ratios remain slightly below 100\%, showing that a finite suffix
space can retain residual ambiguity in extreme collision buckets. The suffix
therefore substantially improves candidate distinguishability but does not
universally guarantee a unique identifier for every node.
\paragraph{Length and codebook capacity.}
Total length four with $K_c=256$ gives the strongest point estimates, but its
differences from total length five are not significant after global Holm
correction. Increasing $K_c$ to 512 also does not improve the point estimates.
The results support length four as an empirical Amazon default, not a
universally optimal configuration.
% \paragraph{Temporal SID stability.}
% The lower block of Table~\ref{tab:sid-mechanism} indicates occasional rather
% than pervasive reassignment: 34.04\% of targets change at least once across
% the full period, but only 4.77\% change at least twice, and fewer than 4.4\%
% change at an average adjacent transition. Because the BERT text embeddings are
% fixed, this measures SID adaptation to graph-snapshot input rather than textual
% semantic drift.
% Snapshot-conditioned and static SIDs have comparable performance: all five
% paired differences are nonsignificant after Holm correction, and the static
% variant has a slightly higher Hits@5 point estimate. Dynamic updating is
% therefore not claimed as a significant accuracy advantage.
\paragraph{Temporal SID stability.}
Across the nine datasets, the mean adjacent-snapshot base-SID change rate
ranges from 0.00\% to 7.35\%. Thus, under every audited snapshot scheme, at
least 92.65\% of candidate nodes retain their base SID across an adjacent
transition on average. IMDB exhibits the largest change rate at 7.35\%,
followed by Amazon Movies and Wikilife at 3.94\%, and Cora at 1.48\%;
the remaining datasets are reported as 0.00\% at two-decimal precision.

These results indicate that snapshot-conditioned SIDs update selectively
rather than changing continually. They remain stable when the quantized
representation is unchanged while retaining the ability to adapt on datasets
that exhibit temporal variation. Because the underlying text embeddings are
fixed in this analysis, the observed changes reflect adaptation to
graph-snapshot inputs and should not be interpreted as direct evidence of
textual semantic drift.
\subsection{Efficiency}
\label{sec:efficiency}

\begin{figure}[t]
    \centering
    \includegraphics[width=\columnwidth]
    {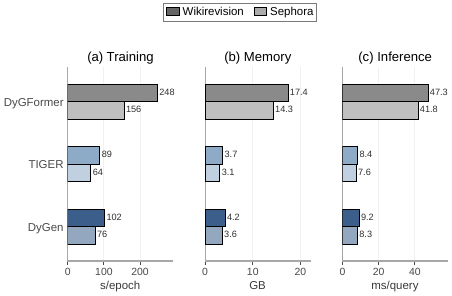}
    \caption{Training and inference efficiency on Wikirevision and Sephora
    under the same hardware configuration. The panels report per-epoch
    training time, peak GPU memory, and per-query inference time at
    $C=5{,}000$, respectively; lower values are better. TGN runs out of
    memory during training, and its inference time is therefore unavailable.
    Exact measurements are shown above the bars.}
    \label{fig:efficiency}
\end{figure}

Figure~\ref{fig:efficiency} compares training and inference efficiency
under the same hardware configuration. Relative to DyGFormer, \method
reduces per-epoch training time by approximately 59\% on Wikirevision and
51\% on Sephora, while reducing peak GPU memory by approximately 76\% and
75\%, respectively. At $C=5{,}000$ candidates, it also reduces per-query
inference time by approximately 80\% on both datasets. Its measured costs
remain close to those of TIGER, which is slightly more efficient on these
benchmarks, whereas TGN exceeds the available GPU memory.

The efficiency gains arise partly because the query context is encoded once
and reused during candidate evaluation. Nevertheless, every prescribed
candidate still requires SID lookup, autoregressive likelihood aggregation,
and ranking. These results therefore demonstrate efficiency at the evaluated
candidate size rather than inference cost that is independent of the number
of candidates.
\vspace{-5pt}
\section{Conclusion}

We presented \method, a candidate-efficient generative framework for dynamic
link prediction on text-attributed graphs. Text-derived SIDs provide a
history-independent representation, while event-level histories refine
prediction as interactions accumulate. Experiments on nine datasets show
broadly competitive overall performance, statistically supported top-rank
advantages on Amazon, and useful source-side inductive generalization across datasets, together with controlled new-user adaptation and target-inductive new-item ranking on Amazon Movies.
The suffix analysis further shows that identifier uniqueness materially
affects ranking. The current results motivate future work on mixed-catalog
calibration, dynamic textual attributes, repeated training seeds, and
continuous-representation controls.

\section*{Ethical Considerations}

This work studies link prediction on benchmark interaction graphs. Such models
can encode popularity, activity, and exposure biases present in historical
data, and cold-start decisions may disproportionately affect newly appearing
users or items. Deployments should therefore audit group- and activity-level
performance, avoid treating predicted links as causal preferences, and apply
privacy-preserving data handling to textual attributes and interaction
histories. Our evaluation uses anonymized benchmark identifiers and reports
aggregate metrics; no attempt is made to infer sensitive attributes.

\paragraph{Generative AI disclosure.}
Generative AI tools were used to assist language editing. All AI-assisted
outputs were reviewed and revised by the authors. The authors take
responsibility for the method, implementation, experiments, and manuscript.

\bibliographystyle{ACM-Reference-Format}
\bibliography{main}

\end{document}